\documentclass[showpacs,preprint,aps]{revtex4}%
\usepackage{amsmath}
\usepackage{graphicx}
\usepackage{amsfonts}
\usepackage{amssymb}%
\begin{document}
\title{Efficient and flexible generation of entangled qudits with cross phase modulation}
\author{Xin Lu Ye}
\email{stardew@hqu.edu.cn}
\author{Qing Lin}
\email{qlin@hqu.edu.cn}
\affiliation{College of Information Science and Engineering, Huaqiao University (Xiamen),
Xiamen 361021, China}


\begin{abstract}
In this paper, we provide a simple but powerful module to generate entangled
qudits. This module assisted with cross-Kerr nonlinearity is available to the
entangled qudits generation with arbitrary dimension, and it could work well
even when the two independent qudits lose the same number of single photons.
Moreover, with the cascade uses of modules, the deterministic generation but
with nonidentical forms of entangled qudits is possible.

\end{abstract}
\maketitle

\section{\bigskip Introduction}

Quantum entanglement plays an important role in the research of quantum
information communication and computing. It has been used widely in various
quantum information tasks, such as quantum teleportation \cite{tel}, quantum
dense coding \cite{coding}, etc. With the development of quantum information,
more and more people turn their attention to high dimension quantum state and
entangled state, since higher dimension means stronger nonlocality and much
more powerful capability for quantum information processing. The application
of qudits, as well as entangled qudits, could increase the security of quantum
cryptography \cite{cry1, cry2, cry3, cry4, cry5, cry6, cry7} and the
efficiency of quantum logic gates \cite{Toffoli1, Toffoli2}. In this sense,
how to create qudits and entangled qudits efficiently is worth discussing. In
optical systems, various schemes are provided, such as orbital angular
momentum entangled qutrits \cite{OAM1, OAM2}, pixel entanglement \cite{pixel1,
pixel2, pixel3}, energy-time entangled qutrits and time-bin entanglement
\cite{time1, time2}, polarization degree of freedom of multi-photon qudits and
entangled qudits \cite{Polar1, Polar2, Polar3, Polar4, Polar5, Polar6, Polar7,
eqdit, bqtrit, Joo, biqutrit}, etc.

Here, we will only focus on the generation of entangled polarization qudits
and we adopt the definition of qudit as $\left\vert j\right\rangle _{n}%
\equiv\left\vert (n-j-1)H,jV\right\rangle $, for $j=0,\cdots,n-1$, where $n$
is the dimension of qudit, and H and V represent the horizontal and vertical
polarizations. In other words, the qudits are represented by the polarizations
of $n-1$ photons in the same spatial and temporal mode. Almost all the
previous proposals of polarization entangled qudits are processed with only
linear optical elements. Therefore we could appreciate the robustness against
decoherence of optical system and the ease of single photon operation
\cite{kok}. However, on the other hand, we have to encounter a serious problem
that the two-particle operations fail most of the time. Unfortunately, this
probability problem is inevitable in linear optical quantum information
processing, including the generation of entangled qudits \cite{eqdit, Joo,
biqutrit}. For example, the success probability of entangled qutrits
generation with linear optical elements is 3/16 in Ref. \cite{Joo} and 1/3 in
Ref. \cite{eqdit}, respectively.

Though various theoretical schemes based on linear optics are considered to be
experimental feasible, however, a realistic quantum information task may
include many qubits and many operations, and then the probability problem will
obviously reduce the possibility of a theoretical scheme. Therefore, a
theoretical scheme with high efficiency or a deterministic scheme is expected,
which is the principle motivation of this paper. Exactly, a deterministic
scheme is possible, if assisted with a new technology, called cross-phase
modulation (XPM) approach. This technology bases on the weak cross-Kerr
nonlinearity and has been used widely in quantum information processing tasks
recently. It enables the deterministic quantum computation with qubits
\cite{xpm1, xpm2, xpm3, computation1, computation2, computation3}, creation of
long-distance entanglement \cite{binghe1, binghe2}, the universal
single-particle operation of qutrit \cite{xqutrit}, etc. Briefly, the XPM
approach bases on the interaction between a Fock state $\left\vert
n\right\rangle $ and a coherent state $\left\vert \alpha\right\rangle $,
resulting in the transformation $\left\vert n\right\rangle \left\vert
\alpha\right\rangle \rightarrow\left\vert n\right\rangle \left\vert \alpha
e^{in\theta}\right\rangle $, where the phase shift of coherent state is
determined by the photon number of the Fock state. Assisted with this XPM
approach, we will show the generation of entangled qudits could be efficient
and flexible. To some extent, this scheme is robust against photon loss, then
the loss of the photon will only result in the reduction of the dimension.
Furthermore, the deterministic generation could be possible with further
operation, though the output forms of entangled qudits are not identical.

The rest of the paper is organized as follows. In Section II, we introduce a
simple module, which could be used to create entangled qutrits from
independent qutrits. In Section III, we will show that this module is
available for entangled qudits generation as well. In order to generate
entangled qutrits or entangled qudits more efficiently, or even
deterministically, we modify the module and then use cascade modules to
achieve this goal in Section IV. Section V is for discussion and conclusion remark.

\section{Generation of entangled qutrits}

\label{sec2}

\begin{figure}[ptb]
\includegraphics[width=7cm]{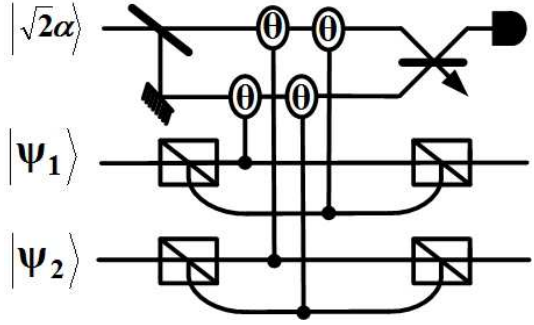}\caption{A simple module and its
application to the entangled qutrits generation. Two independent qutrits are
injected into two PBSs and interacted with two qubus beams $\left\vert
\alpha\right\rangle \left\vert \alpha\right\rangle $ as indicated. After one
more 50:50 BS, an ideal PNND is used to distinguish the vacuum state from
non-vacuum state. If the detection is $n=0$, the entangled qutrits are finally
created. This scheme is suitable for the generation of entangled qudit as
well. For details, see text.}%
\end{figure}

Firstly, we consider the generation of entangled qutrits, which is shown in
Fig.1. This approach enables the transformation from two independent qutrits
to entangled qutrits, and can be applied to the entangled qudits generation as
well (see Sec. III). Suppose the two independent qutrits are initially
prepared as%

\begin{align}
|\psi_{1}\rangle &  =\alpha_{0}|0\rangle_{3}+\alpha_{1}|1\rangle_{3}%
+\alpha_{2}|2\rangle_{3},\nonumber\\
|\psi_{2}\rangle &  =\beta_{0}|0\rangle_{3}+\beta_{1}|1\rangle_{3}+\beta
_{2}|2\rangle_{3},
\end{align}
where $%
{\displaystyle\sum\limits_{i=0}^{2}}
\left\vert \alpha_{i}\right\vert =1$ and $%
{\displaystyle\sum\limits_{i=0}^{2}}
\left\vert \beta_{i}\right\vert =1$. These independent qutrits with some
special forms could be created by higher-order parametric down-conversion
process (two single photons appearing in each of the two output modes)
\cite{bqtrit}, or the Hong-Ou-Mandal (HOM) interference of two single-photon
qubits \cite{HOM}. Alternatively, the independent qutrits with arbitrary forms
could be created by the transformation from spatial encoded photonic qutrits
to biphotonic qutrits as shown in Ref. \cite{xqutrit}.

At first, the two input states are injected into two polarizing beam splitters
(PBSs) respectively. If the qutrit is in state $|0\rangle_{3}$, both photons
will pass through the PBS; if it is in state $|1\rangle_{3}$, one photon will
pass through and the other photon will be reflected; while if it is in state
$|2\rangle_{3}$, both photons will be reflected. After that, we introduced two
quantum bus (qubus) beams $|\alpha\rangle|\alpha\rangle$ and then coupled to
the corresponding photonic modes as depicted in Fig.1. Suppose the XPM phase
shifts induced by the couplings are all $\theta$, and then each mode of the
first qutrit will evolve as follows,
\begin{align}
|0\rangle_{3}|\alpha\rangle|\alpha\rangle &  \rightarrow|0\rangle_{3}%
|\alpha\rangle|\alpha e^{i2\theta}\rangle,\nonumber\\
|1\rangle_{3}|\alpha\rangle|\alpha\rangle &  \rightarrow|1\rangle_{3}|\alpha
e^{i\theta}\rangle|\alpha e^{i\theta}\rangle,\nonumber\\
|2\rangle_{3}|\alpha\rangle|\alpha\rangle &  \rightarrow|2\rangle_{3}|\alpha
e^{i2\theta}\rangle|\alpha\rangle.
\end{align}
Exchanging the orders of output qubus beams in above equations, they are the
evolutions of the second qutrit modes. Then the following state could be
achieved,%
\begin{align}
&  (\alpha_{0}\beta_{0}|0\rangle_{3}|0\rangle_{3}+\alpha_{1}\beta_{1}%
|1\rangle_{3}|1\rangle_{3}+\alpha_{2}\beta_{2}|2\rangle_{3}|2\rangle
_{3})|\alpha e^{i2\theta}\rangle|\alpha e^{i2\theta}\rangle\nonumber\\
&  +\left(  \alpha_{0}\beta_{1}|0\rangle_{3}|1\rangle_{3}+\alpha_{1}\beta
_{2}|1\rangle_{3}|2\rangle_{3}\right)  |\alpha e^{i\theta}\rangle|\alpha
e^{i3\theta}\rangle\nonumber\\
&  +\left(  \alpha_{1}\beta_{0}|1\rangle_{3}|0\rangle_{3}+\alpha_{2}\beta
_{1}|2\rangle_{3}|1\rangle_{3}\right)  |\alpha e^{i3\theta}\rangle|\alpha
e^{i\theta}\rangle\nonumber\\
&  +\alpha_{0}\beta_{2}|0\rangle_{3}|2\rangle_{3}|\alpha\rangle|\alpha
e^{i4\theta}\rangle+\alpha_{2}\beta_{0}|2\rangle_{3}|0\rangle_{3}|\alpha
e^{i4\theta}\rangle|\alpha\rangle.
\end{align}
Next, a 50:50 beam splitter (BS) implementing the transformation $|\alpha
_{1}\rangle|\alpha_{2}\rangle\rightarrow|\frac{\alpha_{1}-\alpha_{2}}{\sqrt
{2}}\rangle|\frac{\alpha_{1}-\alpha_{2}}{\sqrt{2}}\rangle$ will transform the
above state to%
\begin{align}
&  (\alpha_{0}\beta_{0}|0\rangle_{3}|0\rangle_{3}+\alpha_{1}\beta_{1}%
|1\rangle_{3}|1\rangle_{3}+\alpha_{2}\beta_{2}|2\rangle_{3}|2\rangle
_{3})|0\rangle|\sqrt{2}\alpha e^{i2\theta}\rangle\nonumber\\
&  +\left(  \alpha_{0}\beta_{1}|0\rangle_{3}|1\rangle_{3}+\alpha_{1}\beta
_{2}|1\rangle_{3}|2\rangle_{3}\right)  |\alpha_{-}\rangle|\alpha_{+}%
\rangle\nonumber\\
&  +\left(  \alpha_{1}\beta_{0}|1\rangle_{3}|0\rangle_{3}+\alpha_{2}\beta
_{1}|2\rangle_{3}|1\rangle_{3}\right)  |-\alpha_{-}\rangle|\alpha_{+}%
\rangle\nonumber\\
&  +\alpha_{0}\beta_{2}|0\rangle_{3}|2\rangle_{3}|\alpha_{-}^{^{\prime}%
}\rangle|\alpha_{+}^{^{\prime}}\rangle+\alpha_{2}\beta_{0}|2\rangle
_{3}|0\rangle_{3}|-\alpha_{-}^{^{\prime}}\rangle|\alpha_{+}^{^{\prime}}%
\rangle.
\end{align}
where $\left\vert \alpha_{\pm}\right\rangle =|\frac{\alpha e^{i\theta}%
\pm\alpha e^{i3\theta}}{\sqrt{2}}\rangle$ and $\left\vert \alpha_{\pm
}^{^{\prime}}\right\rangle =|\frac{\alpha\pm\alpha e^{i4\theta}}{\sqrt{2}%
}\rangle$.\ Finally, the following desired entangled qutrit (unnormalized),
\begin{equation}
\alpha_{0}\beta_{0}|0\rangle_{3}|0\rangle_{3}+\alpha_{1}\beta_{1}|1\rangle
_{3}|1\rangle_{3}+\alpha_{2}\beta_{2}|2\rangle_{3}|2\rangle_{3},
\end{equation}
could be achieved, if the vacuum state $\left\vert 0\right\rangle $ could be
distinguished from the components $|\pm\alpha_{-}\rangle$ and $|\pm\alpha
_{-}^{^{\prime}}\rangle$. Exactly, an ideal photon number non-resolving
detector (PNND) (on/off detector with quantum efficiency $\eta=1$)\ is
available to complete the discrimination. The success probability is $%
{\displaystyle\sum\limits_{i=0}^{2}}
\left\vert \alpha_{i}\beta_{i}\right\vert ^{2}$. Especially, if $\alpha
_{i}=\beta_{i}=\frac{1}{\sqrt{3}}$ ($i=0,1,2$), the output state is the
maximal entangled qutrit with the success probability $\frac{1}{3}$. The error
probability of this maximal entangled qutrit caused by the overlap of the
coherent components $|0\rangle$, $|\pm\alpha_{-}\rangle$, and $|\pm\alpha
_{-}^{^{\prime}}\rangle$ is%

\begin{equation}
P_{E}=\frac{4}{9}e^{-\left\vert \alpha\right\vert ^{2}\sin^{2}\theta}+\frac
{2}{9}e^{-\left\vert \alpha\right\vert ^{2}\sin^{2}2\theta}.
\end{equation}
Even if $\theta\ll1$ for the weak cross-Kerr nonlinearity, the error
probability will tend to $0$ given $\left\vert \alpha\right\vert \sin\theta
\gg1$ and $\left\vert \alpha\right\vert \sin2\theta\gg1$.

\section{GENERATION of ENTANGLED QUDITS}

\label{sec4}

Actually, the above scheme could be generalized to the generation of entangled
qudits straightly. The generation processes are the same, and suppose the
initial input state is the following product state,%

\begin{equation}%
{\displaystyle\sum\limits_{i=0}^{n-1}}
\alpha_{i}\left\vert i\right\rangle _{n}\otimes%
{\displaystyle\sum\limits_{i=0}^{n-1}}
\beta_{j}\left\vert j\right\rangle _{n}.
\end{equation}
After the input state interacts with the two qubus beams $\left\vert
\alpha\right\rangle \left\vert \alpha\right\rangle $ as depicted in Fig.1, the
following state could be achieved,
\begin{equation}%
{\displaystyle\sum\limits_{i=0}^{n-1}}
\alpha_{i}\beta_{i}\left\vert i\right\rangle _{n}\left\vert i\right\rangle
_{n}\left\vert \alpha e^{in\theta}\right\rangle \left\vert \alpha e^{in\theta
}\right\rangle +\mathcal{C},
\end{equation}
where $\mathcal{C}$ denotes the other cross components that the qubus beams
pick different phase shifts. One more 50:50 BS, associated with an ideal PNND
placing on one of the output modes, will yield the following target state
(unnormalized),
\begin{equation}%
{\displaystyle\sum\limits_{i=0}^{n-1}}
\alpha_{i}\beta_{i}\left\vert i\right\rangle _{n}\left\vert i\right\rangle
_{n}.
\end{equation}
The total success probability is $%
{\displaystyle\sum\limits_{i=0}^{n-1}}
\left\vert \alpha_{i}\beta_{i}\right\vert ^{2}$. If $\alpha_{i}=\beta
_{i}=\frac{1}{\sqrt{n}}$ ($i=0,...,n-1$), the output state is the maximal
entangled qudit $%
{\displaystyle\sum\limits_{i=0}^{n-1}}
\frac{1}{\sqrt{n}}\left\vert i\right\rangle _{n}\left\vert i\right\rangle
_{n}$, and the success probability is $\frac{1}{n}$, reducing linearly with
the increasing of dimension.

All the generation processes are the same for entangled qudit. In other words,
our scheme is suitable for any qudit. The flexibility of the setup in
generating any entangled qudits will be maximized. An arbitrary entangled
qudits could be efficiently generated by a single module in this new approach.
Moreover, to some extent, this approach is robust against photon loss, which
is a realistic problem must be dealt with. If each input state respectively
loses the same number of single photons ($m$), the generation will be still
success, though the level of dimension is reduced to $n-m$, which is another
advantage of our approach.

\section{Deterministic generation of entangled qutrits}

\label{sec3}

\begin{figure}[ptb]
\includegraphics[width=7cm]{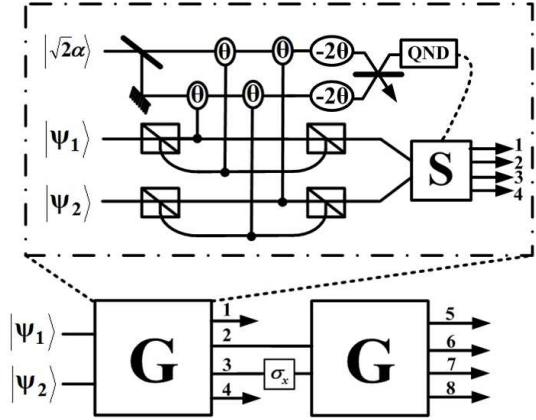}\caption{Deterministic generation of
entangled qutrits with two cascade modified generation modules. In the
modified generation module shown in the dash-dotted line, an additional phase
shift $-2\theta$ is, respectively, applied to two qubus beams after the
interactions. In order to improve the generation efficiency of entangled
qutrits, we replace the ideal PNND by a QND module, which could be used to
realize the projection $\left\vert n\right\rangle \left\langle n\right\vert $,
and the detections are used to control the switch (S) through the classical
feedforward. If the detection is $n=0$, the generation is successful and the
output modes are switched into spatial modes 1 and 4 ; while if the detection
is $n\neq0$, the output modes are switched into the spatial modes 2 and 3 to
be operated further by the second generation module associated with a bit flip
operation placed on spatial mode 3. Finally, the entangled qutrits with other
forms (see Eq. (17) and Eq. (18)) could be achieved in the output modes 5, 8
or 6, 7.}%
\end{figure}

The above scheme is still probabilistic. Since the single photons of the
qutrit or qudit are in the same spatial-temporal mode, then the operations
applied to qutrit or qudit are usually nondeterministically, even assisted
with XPM \cite{xqutrit}. Nevertheless, it is possible to generate the
entangled qutrits and qudits deterministically with further operations, though
the output forms are not identical. Here we use the generation of entangled
qutrits as an example, which is shown in Fig.2.

Similarly, the input states of Eq. (1) interact with two qubus beams as
depicted in the dash-dotted line in Fig.2, and then the state in Eq. (3) could
be achieved. After that, a phase shift $-2\theta$ is respectively applied to
the two qubus beams, which will evolve the state in Eq. (3) to the follows,%
\begin{align}
&  (\alpha_{0}\beta_{0}|0\rangle_{3}|0\rangle_{3}+\alpha_{1}\beta_{1}%
|1\rangle_{3}|1\rangle_{3}+\alpha_{2}\beta_{2}|2\rangle_{3}|2\rangle
_{3})|\alpha\rangle|\alpha\rangle\nonumber\\
&  +\left(  \alpha_{0}\beta_{1}|0\rangle_{3}|1\rangle_{3}+\alpha_{1}\beta
_{2}|1\rangle_{3}|2\rangle_{3}\right)  |\alpha e^{-i\theta}\rangle|\alpha
e^{i\theta}\rangle\nonumber\\
&  +\left(  \alpha_{1}\beta_{0}|1\rangle_{3}|0\rangle_{3}+\alpha_{2}\beta
_{1}|2\rangle_{3}|1\rangle_{3}\right)  |\alpha e^{i\theta}\rangle|\alpha
e^{-i\theta}\rangle\nonumber\\
&  +\alpha_{0}\beta_{2}|0\rangle_{3}|2\rangle_{3}|\alpha e^{-i2\theta}%
\rangle|\alpha e^{i2\theta}\rangle\nonumber\\
&  +\alpha_{2}\beta_{0}|2\rangle_{3}|0\rangle_{3}|\alpha e^{i2\theta}%
\rangle|\alpha e^{-i2\theta}\rangle.
\end{align}
Next, one more 50:50 BS will transform the above state to
\begin{align}
&  (\alpha_{0}\beta_{0}|0\rangle_{3}|0\rangle_{3}+\alpha_{1}\beta_{1}%
|1\rangle_{3}|1\rangle_{3}+\alpha_{2}\beta_{2}|2\rangle_{3}|2\rangle
_{3})|0\rangle|\sqrt{2}\alpha\rangle\nonumber\\
&  +\left(  \alpha_{0}\beta_{1}|0\rangle_{3}|1\rangle_{3}+\alpha_{1}\beta
_{2}|1\rangle_{3}|2\rangle_{3}\right)  |-i\sqrt{2}\alpha\sin\theta
\rangle|\sqrt{2}\alpha\cos\theta\rangle\nonumber\\
&  +\left(  \alpha_{1}\beta_{0}|1\rangle_{3}|0\rangle_{3}+\alpha_{2}\beta
_{1}|2\rangle_{3}|1\rangle_{3}\right)  |i\sqrt{2}\alpha\sin\theta\rangle
|\sqrt{2}\alpha\cos\theta\rangle\nonumber\\
&  +\alpha_{0}\beta_{2}|0\rangle_{3}|2\rangle_{3}|-i\sqrt{2}\alpha\sin
2\theta\rangle|\sqrt{2}\alpha\cos2\theta\rangle\nonumber\\
&  +\alpha_{2}\beta_{0}|2\rangle_{3}|0\rangle_{3}|i\sqrt{2}\alpha\sin
2\theta\rangle|\sqrt{2}\alpha\cos2\theta\rangle.
\end{align}
Now, we replace the ideal PNND in Fig.1 by the projection $\left\vert
n\right\rangle \left\langle n\right\vert $, which could be realized by the
quantum nondemolition detection (QND) module nearly deterministically, even
with the common PNND (quantum efficiency $\eta<1$) \cite{computation2,
computation3, binghe1, binghe2}. The detections will project the above state
into two subspaces. If the detection is $n=0$, the following state could be
achieved (unnormalized),
\begin{equation}
\alpha_{0}\beta_{0}|0\rangle_{3}|0\rangle_{3}+\alpha_{1}\beta_{1}|1\rangle
_{3}|1\rangle_{3}+\alpha_{2}\beta_{2}|2\rangle_{3}|2\rangle_{3}.
\end{equation}
Through the classical feedforward, the above output modes are switched to
spatial modes 1 and 4. While if the detection is $n\neq0$, the output state
could be described as the following unnormalized form,%
\begin{align}
&  c_{1}\left(  \alpha_{0}\beta_{1}|0\rangle_{3}|1\rangle_{3}+\alpha_{1}%
\beta_{2}|1\rangle_{3}|2\rangle_{3}\right)  |\sqrt{2}\gamma\rangle\nonumber\\
&  +c_{2}\left(  \alpha_{1}\beta_{0}|1\rangle_{3}|0\rangle_{3}+\alpha_{2}%
\beta_{1}|2\rangle_{3}|1\rangle_{3}\right)  |\sqrt{2}\gamma\rangle\nonumber\\
&  +\left(  c_{3}\alpha_{0}\beta_{2}|0\rangle_{3}|2\rangle_{3}+c_{4}\alpha
_{2}\beta_{0}|2\rangle_{3}|0\rangle_{3}\right)  |\sqrt{2}\gamma^{^{\prime}%
}\rangle,
\end{align}
where $c_{1}=e^{-2\left\vert \alpha\right\vert ^{2}\sin^{2}\theta}%
\frac{\left(  \sqrt{2}\alpha\sin\theta\right)  ^{n}}{\sqrt{n!}}$,
$c_{2}=e^{in\pi}c_{1}$, $c_{3}=e^{-2\left\vert \alpha\right\vert ^{2}\sin
^{2}2\theta}\frac{\left(  \sqrt{2}\alpha\sin2\theta\right)  ^{n}}{\sqrt{n!}}$,
$c_{4}=e^{in\pi}c_{3}$, $\gamma=\alpha\cos\theta$ and $\gamma^{^{\prime}%
}=\alpha\cos2\theta$. These modes are switched to the spatial modes 2 and 3,
which will be operated further.

It should be noted here that the qubus beam $|\sqrt{2}\gamma\rangle$ or
$|\sqrt{2}\gamma^{^{\prime}}\rangle$ is still strong enough to be recycled as
ancilla, since the XPM phase shift $\theta\ll1$. On the other hand, the
amplitude of the qubus beams is different, but it will not affect the final
result (see below). After the first operation, a bit flip operation
$\sigma_{x}$ is placed on the spatial mode 3, yielding the following
transformations,
\begin{equation}
|0\rangle_{3}\rightarrow|2\rangle_{3},|1\rangle_{3}\rightarrow\left\vert
1\right\rangle _{3},|2\rangle_{3}\rightarrow|0\rangle_{3}.
\end{equation}
Then, the following state could be achieved,%

\begin{align}
&  c_{1}\left(  \alpha_{0}\beta_{1}|0\rangle_{3}|1\rangle_{3}+\alpha_{1}%
\beta_{2}|1\rangle_{3}|0\rangle_{3}\right)  |\sqrt{2}\gamma\rangle\nonumber\\
&  +c_{2}\left(  \alpha_{1}\beta_{0}|1\rangle_{3}|2\rangle_{3}+\alpha_{2}%
\beta_{1}|2\rangle_{3}|1\rangle_{3}\right)  |\sqrt{2}\gamma\rangle\nonumber\\
&  +\left(  c_{3}\alpha_{0}\beta_{2}|0\rangle_{3}|0\rangle_{3}+c_{4}\alpha
_{2}\beta_{0}|2\rangle_{3}|2\rangle_{3}\right)  |\sqrt{2}\gamma^{^{\prime}%
}\rangle,
\end{align}
Similar with the processes from Eq. (2) to Eq. (3) and Eq. (10) to Eq. (11),
we could achieve the following state,%

\begin{align}
&  \left(  c_{1}\alpha_{0}\beta_{1}|0\rangle_{3}|1\rangle_{3}+c_{2}\alpha
_{1}\beta_{0}|1\rangle_{3}|2\rangle_{3}\right)  |-i\sqrt{2}\gamma\sin
\theta\rangle|\sqrt{2}\gamma\cos\theta\rangle\nonumber\\
&  +\left(  c_{1}\alpha_{1}\beta_{2}|1\rangle_{3}|0\rangle_{3}+c_{2}\alpha
_{2}\beta_{1}|2\rangle_{3}|1\rangle_{3}\right)  |i\sqrt{2}\gamma\sin
\theta\rangle|\sqrt{2}\gamma\cos\theta\rangle\nonumber\\
&  +\left(  c_{3}\alpha_{0}\beta_{2}|0\rangle_{3}|0\rangle_{3}+c_{4}\alpha
_{2}\beta_{0}|2\rangle_{3}|2\rangle_{3}\right)  |0\rangle|\sqrt{2}%
\gamma^{^{\prime}}\rangle.
\end{align}
Finally, we use the projection $\left\vert n\right\rangle \left\langle
n\right\vert $ by the QND\ module to detect the first qubus beam. If
$n^{^{\prime}}=0$, we could achieve the following state (unnormalized),
\begin{equation}
c_{3}\alpha_{0}\beta_{2}|0\rangle_{3}|0\rangle_{3}+c_{4}\alpha_{2}\beta
_{0}|2\rangle_{3}|2\rangle_{3},
\end{equation}
from the spatial modes 5 and 8. The unnecessary phase shift $e^{in\pi}$
between the above two coefficients $c_{3}$ and $c_{4}$ could be removed by the
single photon operation $\left(
\begin{array}
[c]{cc}%
1 & \\
& e^{-in\pi/2}%
\end{array}
\right)  $ performed on one of two qutrits. On the other hand, if the result
is $n^{^{\prime}}\neq0$, we could achieve the following state (unnormalized),%

\begin{align}
&  c_{1}\alpha_{0}\beta_{1}|0\rangle_{3}|1\rangle_{3}+c_{2}\alpha_{1}\beta
_{0}|1\rangle_{3}|2\rangle_{3}\nonumber\\
&  +e^{in^{^{\prime}}\pi}\left(  c_{1}\alpha_{1}\beta_{2}|1\rangle
_{3}|0\rangle_{3}+c_{2}\alpha_{2}\beta_{1}|2\rangle_{3}|1\rangle_{3}\right)  ,
\end{align}
from the spatial modes 6 and 7. Similarly, the different phase shifts between
each component could be removed, according to the results $n$ and
$n^{^{\prime}}$. For example, if $n$ is odd and $n^{^{\prime}}$ is even, then
two single photon operations $\left(
\begin{array}
[c]{cc}%
1 & \\
& e^{-i\pi/2}%
\end{array}
\right)  $ respectively performed on the two qutrits could remove these phase shifts.

In the above content, we show two cascade operations, which could project the
input independent qutrits into three forms of entangled qutrits with different
success probability. In other words, the generation of entangled qutrits from
two independent qutrits could be deterministic, though the forms of entangled
qutrits depend on the detections of two QND modules. The situation will be
similar, if we treat the independent qudits following the same method. In the
case of $n$ dimension quantum state, $n-1$ cascade operations will project the
input product qudits into a series of entangled qudits with different forms.

\section{Discussion and CONCLUSION}

\label{sec5}

In this paper, we propose a simple module to generate entangled qudits. This
module is available for any cases and enables deterministic generation by
cascade use. There are other proposals suggested for the entangled qudits
generation with linear optical elements. A heralded two-qutrit entangled state
could be generated through the HOM interferences of two heralded Bell pairs
with the success probability 3/16 \cite{Joo}. Compared with their scheme, our
scheme with one use of module succeeds with the probability 1/3 for maximal
entangled qutrits. Another proposal suggests generating entangled qudits
through the interference of two independent qudits on a PBS, associated with
the ideal photon number resolving QND detection \cite{eqdit}. The success
probability is the same as our scheme. However, the requirement of that QND
detection is not realistic, at least with current experimental technology.
Compared with it, only the ideal PNND is necessary for our scheme, which is
more realistic. In addition, with the theoretical realistic QND module
proposed in Refs. \cite{computation2, computation3, binghe1, binghe2}, the
further operation and the deterministic generation are possible, which is
significantly better than the former works.

Except of the common linear optical elements, the core element of our scheme
is the weak cross-Kerr nonlinearity. Though we treat it in idealized
single-mode picture, it still works well when multi-mode effect is taken into
accounting. It had been theoretically demonstrated that a small conditional
phase shift $\theta\ll1$ with high fidelity could be achieved through the
interaction between continuous-mode photonic pulses (a single photon and a
coherent state) \cite{xc}. In other words, the single-mode approximation in
the weak nonlinearity regime is valid, and then our scheme is feasible with
the current experimental technology.

\begin{acknowledgments}
The authors thank Ru-Bing Yang for helpful suggestions. This work was funded
by National Natural Science Foundation of China (Grant No.11005040), the
Fundamental Research Funds for Huaqiao University (Grant No. JB-SJ1007) and
the Key Discipline Construction Project of Huaqiao University.
\end{acknowledgments}

\end{document}